\documentclass{emulateapj}

\usepackage{natbib}
\usepackage{amsmath}
\usepackage{epsfig}
\newcommand{\multidrizzle}{\texttt{MultiDrizzle}}
\newcommand{\tempbf}{}

\shorttitle{Weak Lensing of A1758}
\shortauthors{Ragozzine \& Clowe}

\begin{document}


\title{Weak Lensing Results of the Merging Cluster A1758\footnotemark[1]}
\footnotetext[1] {Based on observations made with the NASA/ESA Hubble Space Telescope, obtained at the Space Telescope Science Institute, which is operated by the Association of Universities for Research in Astronomy, Inc., under NASA contract NAS 5-26555, under program 11194.  Also based on data collected at the Subaru Telescope and partly obtained from SMOKA, which is operated by the National Astronomical Observatory of Japan.}
\author{B. Ragozzine, D. Clowe \footnotemark[2]}
\footnotetext[2]{Sloan Fellow}
\affil{Department of Physics and Astronomy, Astrophysical Institute, Ohio University, Athens, OH 45701}
\email{ragoz@phy.ohiou.edu}

\author{M. Markevitch}
\affil{NASA GSFC, Code 662, Greenbelt, MD 20771}

\author{A. H. Gonzalez}
\affil{Department of Astronomy, University of Florida, Gainesville, FL 32611-2055}

\and

\author{M. Brada{\v c}}
\affil{Department of Physics, University of California Davis, Davis, CA 95616}


\begin{abstract}

Here we present the weak lensing results of A1758, which is known to consist of four subclusters undergoing two separate mergers, A1758N and A1758S.  Weak lensing results of A1758N agree with previous weak lensing results of clusters 1E0657-558 (Bullet cluster) and MACS J0025.4-1222, whose X-ray gas components were found to be largely separated from their clusters' gravitational potentials.  A1758N has a geometry that is different from previously published mergers in that one of its X-ray peaks overlays the corresponding gravitational potential and the other X-ray peak is well separated from its cluster's gravitational potential.  The weak lensing mass peaks of the two northern clusters are separated at the $2.5\,\sigma$ level.  We estimate the combined mass of the clusters in A1758N to be $2.2\pm0.5\times10^{15} M_\odot$ and $r_{200}=2300_{-130}^{+100}$ kpc.  We also detect seven strong lensing candidates, two of which may provide information that would improve the mass measurements of A1758N.

\end{abstract}




\section{Introduction}

Current cosmological models show that baryonic matter comprises 4.6\% of the total mass distribution of the universe and that dark matter comprises 23\% \citep{komatsu10}, thus the dark matter component is approximately five times more abundant than baryonic matter.  Baryonic matter follows the distribution of dark matter in the universe and in order to test its existence the baryonic matter must be separated from the dark matter component, but this is problematic because so far it has only been detected during the merging process of galaxy clusters.

Galaxy clusters are permeated by X-ray gas, which makes up approximately 90\% of the baryonic matter in the cluster (e.g. \cite{david90, gonzalez07}).  As merging clusters interact, the X-ray gas clouds are slowed through ram pressure while the galaxies and the collisionless dark matter continue nearly uninterrupted \citep{furlanetto02}.  Thus, merging clusters can provide a unique view where, given the proper configuration, the dark matter component is separated from the X-ray gas.  The presence of dark matter can be inferred by showing an offset of the cluster's overall gravitational potential from its massive X-ray plasma via gravitational lensing analysis.  Gravitational lensing, which is the bending of light of background galaxies as it passes near a massive structure, is independent of the type of matter present and thus indicates the overall gravitational potential of the cluster (see review by \cite{bartelmann01}).  

An offset between the X-ray gas distribution and the mass inferred from gravitational lensing has been detected in the Bullet cluster \citep{clowe06} and MACS J0025.4-1222 \citep{bradac08}.  Alternate gravity models, which replace the need for dark matter with gravity being stronger than Newtonian on Mpc scales (e.g. \cite{milgrom83, bekenstein04, moffat06}), do not completely model the mass of the merging clusters with the optical and X-ray components alone.  Instead, alternate gravity models need some smaller amount of dark matter compared to cold dark matter models, but still require at least twice the amount of matter in some dark form as is present in the baryonic component \citep{brownstein07, feix08}.  Alternate gravity models that recover the mass distribution of the Bullet cluster can likely model MACS J0025.4-1222 because of their similar geometries (the X-ray peaks are between the lensing peaks).  However, given a system with the same mass and different geometry, alternate gravity plus dark matter models get a different answer than general relativity \citep{angus06}.  Results are needed from additional merging clusters with different geometries in order to determine which model is correct.  Additionally, merging clusters are one of the few ways to measure the self-interaction cross section of dark matter.

A1758, which is composed of four clusters, is undergoing two separate mergers.  Evidence that the two clusters in A1758N (both are 7 keV clusters) are merging includes a projected separation of 800 kpc between its two clusters, a disturbed X-ray gas distribution, and a separation of the X-ray gas from one of the clusters.  Evidence that the less massive clusters in A1758S (both 5 keV clusters) are undergoing a merger is that there is very little projected separation between the clusters and it has a disturbed X-ray morphology \citep{david04}.  There is no evidence of a merger between A1758N and A1758S in the X-ray signature and they have a projected separation of 2.0 Mpc \citep{david04, okabe08}.  A1758N introduces a new geometry that is different from previously published mergers.  One weak lensing peak overlaps an X-ray peak while the other weak lensing peak is clearly separated from the X-ray component \citep{david04, okabe08}.  The dynamical state of A1758N given in \cite{david04} is that the southeast cluster is moving to the southeast and the northwest cluster is moving to the north.  \cite{david04} conclude that the clusters in A1758S appear to be at the point of closest approach, but could be a projection effect.  However, \cite{haines09} argue against A1758S being at closest approach based on its lower brightness in the core.  There is very little redshift information for A1758 ($z=0.279$).  \cite{david04} show that A1758N and A1758S have a velocity difference of less than 2100 km s$^{-1}$, which is consistent with infall velocity at the redshifts they fit from spectral analysis of the X-ray gas.

The previous lensing analysis by \cite{okabe08} was done with two band filters.  This analysis improves upon their work by using deeper images to get better shape information of background galaxies and uses a third filter, which allows for better separation of cluster galaxies from the background.  As a result, this work has a more detailed weak lensing shear signal and mass reconstruction of A1758.

Throughout this paper, we use the cosmological parameters $H_0=70$ km sec$^{-1}$ Mpc$^{-1}$, $\Omega_M=0.27$ and $\Omega_\Lambda=0.73$ \citep{komatsu10}.

\section{Observations}\label{sec:observations}

For our analysis, we used observations of A1758 taken with two observatories, with the goal of obtaining a broad ground-based field of view (Subaru) for a weak lensing analysis of the cluster with improved spaced-based imaging (Hubble) of the cluster cores.  Here we describe the images we acquired and the reduction process used to prepare just the ground-based  images for the weak lensing measurements.  In the end, the Hubble images were not adequate for getting shape measurements, but were used to {\tempbf identify strong lensing candidates}.  The weak lensing analysis is described in Section \ref {sec:analysis}.

\subsection{Subaru}\label{subsec:subaru}

Observations of A1758 were carried out on the Subaru telescope's Suprime Cam \citep{miyazaki02}, which is a wide-field camera whose field of view is $34^\prime \times 27^\prime$.  Images in $B$ and $V$ passbands were taken on April 7, 2008 using time acquired from the Gemini time-share program.  The weather for the night was photometric.  Each of our images were 300-sec exposures, which we chose in order to minimize the number of saturated stars while keeping CCD readout overhead to an acceptable level.  In total we took 6 images in $B$ and 16 images in $V$, giving 30 and 80 minutes total integration time respectively.  The telescope was dithered by 20$\arcsec$ after each image was taken to fill in chip gaps and allow for night sky flats to be created.  The $B$ images had a range of $0\farcs80-0\farcs91$ seeing and a typical value (of the combined images) of $0\farcs85$.  The $V$ images ranged from $0\farcs76-0\farcs90$ and had a typical value of $0\farcs81$.  

To get a third passband, we used archived Subaru Suprime Cam $R_c$ images, downloaded from SMOKA \citep{baba02} which were taken on July 17, 2007.  We used 6 images, each with a 500-sec exposure time, giving a total exposure time of 50 minutes.  The images in $R_c$ were taken with a different rotation angle, therefore the field of view of the $R_c$ band did not completely overlap that of the $B$ and $V$ images, but did cover both the A1758N and A1758S regions.  The seeing in the $R_c$ images ranged from $0\farcs83-0\farcs92$ and had a typical value of $0\farcs87$. After rotating the $R_c$ image into the orientation of the $B$ and $V$ images the typical seeing increased to $0\farcs90$.

Image processing was as follows.  A master bias file was created for each of the ten chips by averaging pixel values within $3\,\sigma$ of the initial mean.  These master bias files were subtracted from the science frames and then a linear fit was made of the overscan strip to subtract any residual bias changes.  A master flat was first created for each chip from the bias subtracted dome flat images and each chip was smoothed and visually analyzed for steep gradients and dust or other undesirable chip features.  Typical pixel values were kept in these normalized flats in the range of $0.86-1.03$ and the remaining pixels were marked bad.  The four corner chips, however, had a steep gradient in the far corners of the CCD array and pixel values were kept in the range of $0.75-1.07$ to reject the worst parts of the flats while keeping as much of each corner chip as possible.  The science frames were divided by the master flat of the corresponding chip.  Another set of flats was created from the flattened science frames and having multiple pointings allowed for excluding individual chips with saturated stars.  The science frames were then divided by these new flats according to chip.

The sky background was subtracted from each file by making a smoothed image of all local minima within $5\,\sigma$ of the mode of pixel values in order to avoid minima in saturated stars.  The chip images were scaled so that they had equal gain and the same sky levels.

The brightest stars and galaxies in A1758 were then aligned with 633 matching objects in the USNO catalog by fitting a $1^{st}$ order polynomial (3 free parameters per chip: $dx$, $dy$, and rotation) to get the ten chip planes into the detector plane.  We kept one chip fixed and found a solution to the other nine chips in relation to it (27 free parameters, which we kept constant for all images).  We simultaneously created a $7^{th}$ order 2D polynomial solution for each image from the detector plane to the sky (36 free parameters per image) with 43,119 total stars (stars per image ranged from $521-699$) from 75 images taken on April 7, 2008, which is much larger than the $(36\times75)+27$ free parameters needed for the solution.  We created a separate $7^{th}$ order solution for the $R_c$ images taken on July 17, 2007 with an average of 695 stars per image.  Our mapping to the USNO catalogs was accurate to $0\farcs43$ (median), consistent with the accuracy of the catalog itself.  The consistency of the stars between images in this mapping process was accurate to within $0\farcs034$ (median) and there were no large scale gradients in the residual offsets in these measurements.

Standard stars taken on April 7, 2008 include the fields SA98, SA104, and two pointings of SA110 (the fields taken on July 17, 2007 include SA104 and SA107) and were reduced in the same manner as above.  Standards that lay on bad pixels were discarded if the pixels lay within several half-light radii of the centroid.  The zeropoint magnitudes were calculated as a function of color and airmass with rms variance in $B$, $V$ and $R$ of 0.031, 0.014 and 0.019, respectively, and is the primary error of our photometry.  There were no large scale gradients in the zeropoint levels or errors by position.

\subsection{Hubble Space Telescope}\label{subsec:hst}

We obtained observations of the A1758N substructure field using a 28 pointing mosaic with the Wide Field Planetary Camera 2 (WFPC2) in the F606W filter in cycle 16, converting the original 3 passband, 4 pointing ACS proposal to a single passband WFPC2 mosaic after the ACS shutdown the night after the cycle 16 proposals were due.  The observations were taken between October 23rd and 28th, 2007.  The mosaic was arranged in two layers, a 16 pointing first layer and a 12 pointing second layer.  The pointings were taken with the same approximate roll angle, with the second layer pointings offset by roughly half a WFPC2 chip in size ($\sim 40\arcsec$).   The offset was done in order to attempt to identify the charge transfer inefficiency (CTI) pattern in the ccds and correct it as part of image reduction in a similar manner to how the CTI in the Advanced Camera for Surveys has been corrected \citep{massey10}.  The WFPC2 CTI was so bad, however, that we were unable to create a suitable correction scheme that would remove the CTI from being the dominant systematic error in galaxy shape measurements.

To process the images we used the standard calibration data and a modified version of custom software from the HAGGLes project (Marshall et al. 2011, in preparation), which is based upon \multidrizzle~\citep{koekemoer02}. Specifically, we adapted the code to be compatible with WFPC2 data and disabled iterative astrometric matching due to the limited WFPC2 field-of-view.  The results are the same as using \multidrizzle. To create a final mosaic we utilized the {\it SCAMP} v1.4.6 and {\it SWARP} v2.17.1 packages \citep{bertin02}.  The final mosaic has a mean internal astrometric uncertainty of $0\farcs021$.  Due to the lack of a CTI correction, we found that we could not trust the weak lensing measurements enough to believe any substructures in the HST data, and so have used only the Subaru images for the weak lensing analysis below.  We use the HST images in the strong lens identification section (Section \ref{sec:stronglens}).

\section{Weak Lensing Analysis}\label{sec:analysis}

\begin{figure*}
	\epsscale{1.0}
	\plotone{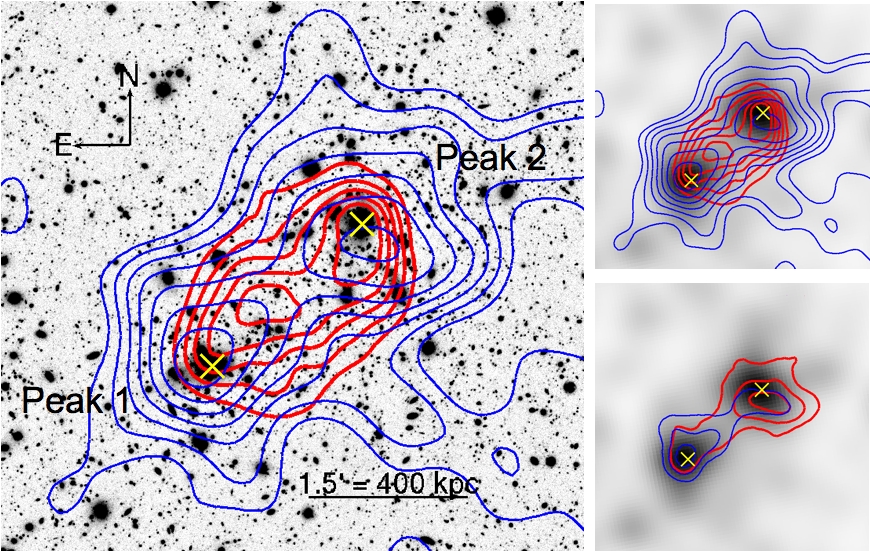}
	\caption{The left panel is a gray scale image of the 80-minute $V$-band image of the A1758N cluster merger ($z=0.279$); the two BCGs are separated by $\sim 3\arcmin$ (800 kpc) on the sky; there could also be a radial component.  The blue contours represent the weak lensing mass reconstruction made from a background galaxy density of 24.0 galaxies/arcmin$^2$.  The outer blue contour begins at $\kappa = 0.07$ and each contour increases in steps of 0.045 up to $\kappa = 0.34$.  The red contours follow the X-ray gas mass based on a 41.5 ksec Chandra exposure.  The NW cluster's BCG aligns with the X-ray gas and the weak lensing peak.  The SE cluster's BCG and weak lensing peak are well separated from the X-ray gas, which has a bright peak near the midpoint of the two weak lensing peaks.  The two small panels show the smoothed red sequence cluster light, which has been weighted by flux.  The contours in the upper right panel are the same contours as in the left panel.  The contours in the lower right panel show the $1-3~\sigma$ errors on the centroid for the weak lensing analysis.}
	\label {fig:A1758}
\end{figure*}

After we processed the Subaru images and combined them by filter, we used SExtractor \citep{bertin96} on the final images to create a catalog of objects and measure their brightness.  Then we used a modified version of the imcat software\footnote{http://www.ifa.hawaii.edu/$\sim$kaiser/imcat/} to measure size, shape, and significance of the objects in the catalog.  The resultant catalogs were purged of objects with less than $10\,\sigma$ (to not dilute the shear signal), with an extremely high ellipticity (0.6, to reject cosmic rays that are nearly parallel to the detector), or that had any bad pixels (which would bias the centroid, flux and shape measurements).  

We determined the PSF by plotting all objects in radius vs.\ magnitude space and selecting those objects with constant radius ($\sim 2.0$ pixels) and $V$ as faint as $\sim 24$ mag.  These objects are mainly stars, but may include some small, faint galaxies.  These point-like (or very small) objects have shapes that are entirely (or heavily) influenced by the PSF and provide us a way to model it in each image.  We want isolated objects so that close neighbors do not negatively influence our PSF model.  After removing objects that had neighbors within 30 pixels ($\sim 15$ half-light radii), the remaining objects were modeled using a $10^{th}$ order 2D polynomial to iteratively fit the objects and reject anomalies.  The remaining stars were then used to calculate the PSF.

In order to isolate the background galaxies in the full catalog, we removed all objects  from the catalog that had a smaller half-light radius than the largest stars used to create the PSF model ($\sim$ 2.2 pixels).  Objects were also removed if they had an unusually large maximum pixel value compared to their overall brightness, which indicates something out of the ordinary like a cosmic ray or a faulty pixel.  The objects that remain in the catalog are predominantly the cluster, foreground, and background galaxies.

In order to determine the cluster and foreground galaxies, we plotted simulated galaxies from a catalog provided by R. Pell{\'o} (priv.~comm.), which is based on spectral energy distribution templates of the hyperz photometric redshift code \citep{bolzonella00}.  We plotted galaxies with $z < 0.3$ in $B-V$ vs.~$V-R$ and compared them to the galaxies within $3\arcmin$ of the two BCGs in A1758N.  We then applied a box cut around the knot of red sequence galaxies and the trail of bluer (spiral) galaxies.  This box cut was applied to our catalog to remove the cluster and foreground galaxies, leaving just the background galaxies.

Next, we discarded any background objects if the sky gradient around them was too steep, meaning they were too close to a bright star or large galaxy and their shapes could be biased.  Finally, we discarded galaxies dimmer than $V \sim 27$ and $R\sim26$, because they are heavily influenced by the PSF and their true shapes are not well measured.

We measured the shear of the resultant catalog of background galaxies by calculating the second moment of their surface brightness and removed the modeled PSF using the KSB method \citep{kaiser95, luppino97}, which removes the ellipticity of the PSF from the shear of each object.

Our weak lensing analysis of A1758 was applied to the STEP2 data set \citep{massey07} in order to determine the amount of shear recovered in this analysis.  The data sets psfA and psfC are most like the PSF conditions of the Subaru telescope where the seeing is in the range of $0\farcs6-0\farcs8$.  The \citet{luppino97} method has a well known bias that under measures the shear and, based on our analysis of the STEP2 data, we determined that we recovered 0.893 and 0.875 of the shear in the $V$ and $R_c$ images.  We divided the shear of each object by the respective value in order to correct for this bias.  

{\tempbf To justify our comparison to STEP2, we note that the similarities between our images and those simulations include pixelation, correlated noise, realistic galaxy shapes and a realistic PSF.  The only significant differences are the potential changes caused by cosmic ray clipping routines, the exact form of the correlated noise and PSF shapes, and that the PSF was constant in STEP2.  In our analysis and in STEP2 results, correlated noise does not appear to significantly affect shape measurements performed using KSB-type PSF correction, and any potential effect would be significantly smaller than the level of random noise in individual cluster mass reconstructions.  Our co-adding routines were designed to err on the side of leaving cosmic rays in the center of bright objects to avoid pixel clipping that can alter object shapes, particularly those of stars, and have been tested on simulations. All of these potential effects have been shown to change the shears in both STEP2 and similar simulations at the few percent level, which is small compared to the random errors in the measurements presented in this paper.}

We then placed galaxies into bin regions based on size and significance and weighted each object according to $1/rms$ of the shear of the bin.  The shear of objects with the lowest SN are thus given the least weight, $w$.  We reconstructed the mass of A1758 by applying a Fourier transform to the shear from the $V$ and $R_c$ catalogs separately and found that the two maps were in good agreement with each other and with the BCGs.   We created our final convergence map by using background galaxies that appeared in both the $V$ and $R_c$ catalogs, calculated the final reduced shear of each galaxy as $g=(w_v^2g_v+w_r^2g_r)/(w_v^2+w_r^2)$ and the final weights were added in quadrature; see the convergence map in Figure \ref{fig:A1758}.  We computed $\Sigma_{crit} = 3.21\times10^9$ $M_\odot/$kpc$^2$ by taking the COSMOS photo-z catalog of \cite{ilbert09}, applying the same color and magnitude cuts as we used on our background galaxy catalog and taking the harmonic mean, which was $z_{bg} = 0.784$.

\subsection{X-ray Mass}

We have created an approximate projected gas mass map using the Chandra X-ray image in the 0.8--3.5 keV energy band extracted from the 41 ks observation (OBSID 2213) analyzed in \cite{david04}. For a hot cluster such as A1758 ($T_e=9.0$ keV, David \& Kempner), the X-ray emissivity at photon energies $E\ll T_e$\/ depends very weakly on gas temperature (and its likely variations across the cluster).  Chandra has a peak of sensitivity at $E\simeq 1$ keV, which makes the X-ray surface brightness in our energy band a good representation of the projected X-ray emission measure, $EM\propto n_e^2$.

To convert the projected emission measure to the gas mass requires knowledge of the three-dimensional cluster geometry.  Unlike the Bullet cluster \citep{clowe06, markevitch02}, which appears to have a simple geometry, the geometry of A1758 is more complicated. In the absence of any information or plausible assumptions on gas distribution along the line of sight, an approach frequently used to get a first-order approximation of a gas mass map is simply to take a square root of the X-ray brightness.  Since on large scales, clusters have spherical, centrally concentrated density distributions, here we use a slightly more accurate approximation.  We first fitted a spherically-symmetric $\beta$-model to the X-ray radial brightness profile and created a projected gas mass map that corresponds to that model.  This mass map was then multiplied by a correction factor $(S_X/S_\beta)^{1/2}$, where $S_X$ is the real X-ray image and $S_\beta$ is the $\beta$-model image. To normalize the gas mass, we used an $M_{\rm gas}-T$\/ relation from Vikhlinin et al.\ (2009; technically, we combined their $M_{\rm tot}-T$\/ and $f_{\rm gas}-T$\ relations, which were derived from the Chandra gas masses and X-ray temperatures).  Though the \cite{vikhlinin09} $M_{\rm gas}-T$\/ relation is derived for relaxed clusters, hydrodynamic simulations indicate that it should not be very different for mergers \citep{nagai07}. For the A1758 temperature ($T=9.0$ keV) and $z$, the relation gives $M_{\rm gas}=1.1 \pm 0.28 \times 10^{14} M_\odot$ in a sphere of radius $r_{500}=1.16$ Mpc. We normalized the resulting gas mass map to have the projected mass in an aperture of the radius $0.65 r_{500}$ (which corresponds to the Chandra field of view) equal to that of the best-fit $\beta$-model with the above mass within the $r_{500}$ sphere.  The resulting gas mass map is shown in Figure \ref{fig:A1758}.

{\tempbf Simulations, e.g., by Kravtsov et
al. (2006) and Rasia et al. (2011), indicate that even the clusters in
the middle of a major merger, such as A1758, follow the $M-T$ relation
with a scatter of about 20--25\% along the mass axis (see, e.g., Figure 6 in Rasia et al. for quantities within $r_{500}$), and the $M_{\rm
gas}-M_{\rm tot}$\/ relation is even tighter. Other errors in our
analysis should be smaller and we have assigned a conservative 25\% error (68\%
confidence) to the gas masses.}

\section{Results of A1758}

Our WL mass reconstruction of A1758N was created by using 20,919 background galaxies that appear in both the $V$ and $R_c$ images.  The overlapping area between the $V$ image and the rotated $R_c$ image is $\sim 791$ arcmin$^{2}$ and results in a density of 24.0 galaxies/arcmin$^{2}$ that is used for shear measurements.  We created the convergence map (see Figure \ref{fig:A1758}) using the KS93 method \citep{kaiser93}, which takes the shear catalog {\tempbf(described in Section \ref{sec:analysis}) and directly inverts it using a Fourier transform.  This produces a 2D projection of the mass along our line of sight.  We smoothed the mass map enough to smooth over small noise features, but not so much that our weak lensing peaks would blend together.}  The BCGs of both the SE and NW clusters line up well with the weak lensing peaks in our analysis, which are labeled Peak 1 and Peak 2, respectively.  

\subsection{Cluster Light}

\begin{deluxetable*}{lccccc} 
\tablecolumns{6}
\tablecaption{Mass\tablenotemark{a} \label{table:mass} } 
\tablehead{   
  \colhead{} & 
  \colhead{$M_{\rm ap}$} & 
  \colhead{$L$} & 
  \colhead{$M_{\rm ap}/L$} & 
  \colhead{$M_{\rm gas}$} & 
  \colhead{($M_{\rm ap}-M_{\rm gas})/L$} \\
  \colhead{} & 
  \colhead{$(10^{13}~M_\odot$)} & 
  \colhead{$(10^{11}~L_\odot)$} & 
  \colhead{$(M_\odot/L_\odot)$} & 
  \colhead{$(10^{12}~M_\odot)$} & 
  \colhead{$(M_\odot/L_\odot)$}
}
\startdata
Peak 1		& 8.60		$\pm$ 1.0	& 4.08		& 210.7 $\pm$ 24.1		& 7.39 $\pm$ 1.85	& 192.6 $\pm$ 22.0 \\
Peak 2		& 8.58		$\pm$ 1.0		& 4.28		& 200.6 $\pm$ 23.3		& 7.64 $\pm$ 1.91	& 182.7 $\pm$ 21.2 \\
Midpoint		& 7.51		$\pm$ 1.0	& 2.09		& 358.5 $\pm$ 46.9		& 8.33 $\pm$ 2.08	& 318.7 $\pm$ 41.7 \\
Left			& 3.30		$\pm$ 1.1		& 1.07		& 306.8 $\pm$ 106.1	& N/A	& N/A  \\
Right			& 4.97		$\pm$ 1.1		& 1.45		& 341.0 $\pm$ 72.0		& N/A	& N/A  \\
Peak 3		& 5.72		$\pm$ 1.1		& 2.74		& 208.7 $\pm$ 40.4		& N/A	& N/A \\
Peak 4		& 5.54		$\pm$ 1.1		& 2.20		& 251.5 $\pm$ 50.1		& N/A	& N/A \\
Random 1	& 2.48		$\pm$ 1.1		& 0.33		& - 						& -		& - \\
Random 2	&-0.280	$\pm$ 1.2		& 0.14		& -							& -		& - \\
Random 3	& 0.197	$\pm$ 1.3		& 0.14		& -							& -		& - \\
Random 4	& 0.448	$\pm$ 1.3		& 0.082	& -							& -		& - \\
\enddata

\tablenotetext{a}{Mass measurements are based on identical, non-overlapping circular apertures centered on the $\kappa$ peaks (and other regions) from the weak lensing analysis.  The $\kappa$ within the aperture radius $r= 0\arcmin.663=169.7$ kpc was converted to mass (see Section \ref{subsec:mass} for details).}

\end{deluxetable*}

Red sequence cluster galaxies were selected by first finding the red sequence of the objects within $3\arcmin$ of the two BCGs in a color-magnitude diagram and then selecting all objects in the catalog with the same color-magnitude selection.  These cluster galaxies were compared both by flux and by number density; the results were consistent with each other.  The side panels in Figure \ref{fig:A1758} show the flux weighted cluster light.  The cluster light image was also used to select random locations away from the cluster galaxies for noise measurements of the mass in the bootstrap resamplings (see section \ref{subsec:mass}).

\subsection{Errors on the Centroid and Cluster Core Masses}\label{subsec:mass}

The errors on the centroids for Peaks 1 and 2 were calculated by performing $\sim 3\times10^5$ bootstrap resamplings of the background galaxies, recreating the convergence map, and plotting the nearest {\tempbf centroid (calculated with imcat's findpeaks routine)} with each resampling (see lower right panel of Figure \ref {fig:A1758}).  The results are the green and red contours, which show one, two, and three standard deviations of the centroids of Peaks 1 and 2.  Peak 1 (2) was found to the NW (SE) of the midpoint 1.19\% (2.02\%) of the time.  Figure 1 (lower right plot) shows that the separation of Peaks 1 and 2 is significant at the level of $\sim 2.5\, \sigma$. {\tempbf Bootstrapping does not detect systematic errors, such as imparting a false shape due to stellar shape.  To account for this systematic, we measured the error that would occur if the stellar shape were not properly removed and thus affect the centroid measurement.  If left uncorrected, the mass reconstruction based solely on the stellar shear field at Peak 1 (2) would shift the centroid by a maximum amount of $2\arcsec$ ($6\arcsec$).  This is a smaller shift than the 68\% contour levels in Figure \ref {fig:A1758}.  The KSB method, however, removes the PSF down to a few percent \citep{massey07}.  Effects for the remaining unsubtracted portion from the stellar shear field are, therefore, negligible.}

The resampled peaks of both Peak 1 and Peak 2 that were found past the midpoint (that coincides with an X-ray peak) were analyzed to determine why the peak finder would completely miss one peak and find it at the location of the other peak.  The resampled convergence maps always showed a significant amount of mass at both locations, but in these cases the mass signal steadily increased from one peak, across the intermediate region, and continued increasing toward the other peak.  Finding the centroid away from the expected location was due to the linear peak finder and {\it not} due to a lack of mass in the weak lensing signal at the location of the peak in question.  The noise in these resamplings occurred in such a way as to create a smooth mass increase away from the peak being calculated.

To quantify the amount of mass in these regions we averaged the reconstructed mass map pixel values (in units of $\kappa$) in identical, non-overlapping circular apertures with as large of a radius as possible.  The radius used for the apertures was $0\farcm663$, or 169.7 kpc at the redshift of A1758.  The mean $\kappa$ within each cutout was converted to mass by multiplying by $\Sigma_{crit}$ and the area of the aperture.

When Peaks 1 and 2 were found on the opposite side of the midpoint, the average mass of these bootstrap resamplings were low by $\sim 2\, \sigma$.  At the same time, the midpoint was $\sim 0.5\,\sigma$ higher than its average; in this way the peak finder wandered up the mass structure away from the original region.

The statistics of these points of interest in A1758N, as well as four other randomly selected regions (away from the red cluster light), show that they are Gaussian distributed.  The distributions of Peaks 1, 2, and the midpoint and that of the Random regions are in good agreement with each other.  In units of $10^{13}\,M_\odot$, the mean values of the three peaks are: $m_{Peak 1} = 8.60 \pm 1.0$, $m_{Peak 2} = 8.58 \pm 1.0$, $m_{midpoint} = 7.51 \pm 1.0$ (see Table \ref{table:mass}).  The significance of these masses are $8.7\, \sigma$, $8.6\, \sigma$, and $7.6\, \sigma$.  

The mean mass at Peak 2 and the midpoint are only different by $\sim 1\, \sigma$, so we tested the mass results for linear correlation.  Using $3.76\times10^4$ bootstrap resamplings, we found that the mass in Peak 1 (2) was correlated with the mass in the midpoint with a coefficient of 0.309 (0.211), or $59.9\,\sigma$ ($40.9\,\sigma$) above zero correlation.  Peaks 1 and 2 were correlated with a coefficient of 0.0140, or $2.72\,\sigma$ above zero correlation.  The mass in Peak 1 (2) was larger than that in the midpoint 82.8\% (80.8\%) of the time and there was more mass in Peak 1 compared to Peak 2 50.5\% of the time.

These correlations are likely due in part to smoothing used in the mass map and in part because the large aperture cutouts nearly touch; the mass in each resampling moves with the centroid, pushing mass out of the apertures---sometimes toward the midpoint.  We also compared the mass within smaller apertures (half the radius, 84.85 kpc).  The linear correlation coefficients between Peak 1 (2) and the midpoint were 0.191 (0.130), or $37.1\,\sigma$ ($25.3\,\sigma$) above zero correlation.  Peaks 1 and 2 were correlated with a coefficient of 0.0188, or $3.6\,\sigma$ above zero correlation.  With this smaller aperture, the mass in Peak 1 (2) was larger than the mass in the midpoint 87.6\% (92.9\%) of time and Peak 1 was larger than Peak 2 33.0\% of the time.

\subsection{Cluster Virial Masses}\label{subsec:virial}

Now that we have determined that there is a considerable amount of mass at Peaks 1 and 2, the next measurement of interest is to determine the masses of the clusters in A1758N.  Fitting NFW profiles to these clusters depends on the input parameters for the inner and outer radii of the fit.  Because the weak lensing approximation breaks down in the strong lensing region, the inner radius should be larger than the strong lensing radius.  No strong lensing has yet been confirmed in this system so we used a value of 100 kpc for the inner radius.  The outer radius that we used was 3350 kpc, which is the distance from the clusters in A1758N to the nearest edge of our image.  This outer radius is much larger than $r_{200}$ of both clusters (shown below).

We also allowed our color-color selections to vary (see Section \ref{sec:analysis}) because the Subaru filters differ slightly from those in the R. Pell{\'o} catalog and because there may be systematic errors in our zeropoints.  We maximized the significance of our cluster fitting by allowing the redder side of our box cut to increase redwards by 0.5 mag; other shifts in the box edges did not improve the significance of the clusters.  

We attempt to simultaneously fit the individual clusters to obtain their masses, but find that with the current data we cannot robustly extract mass measurements.  Initially, we fit all four clusters simultaneously (two in A1758N and two in A1758S) with NFW profiles.  However, the best fit to one of the weak lensing peaks in A1758S was not a well-behaved cluster, but had a very large ($> 2.0$ Mpc) $r_{200}$ with extremely low ($c<10^{-3}$) concentration.  We also tried to fit each cluster in A1758S separately (with and without the clusters in A1758N), but the fitting behavior in A1758S persisted in these extreme values for radius and concentration.  In order to determine the mass in the cluster members of A1758N from NFW profiles, we need to decide what to do with the members in A1758S.  The effect of including both southern members is to remove mass from the northern members and place it in an extended (and unphysical, $m>2\times10^{16}M_{\odot}$) cluster.  Excluding both southern members greatly increases the radius and mass of the clusters in A1758N.  The results of fitting just the northern clusters simultaneously produces different mass results than when we fit them one at a time.  When fit together, one northern member is five times larger than the other.  When fit separately, the northern clusters are approximately the same size.  Whether we fit them together or separately, the NFW profiles of both of these methods have approximately the same significance.  Our conclusion is that the northern peaks are poorly fit by the classic NFW profile and we do not trust this method of measuring the cluster component masses.  For comparison, however, fitting two clusters in A1758N (with a single cluster in A1758S) result in $r_{200}$ values (and concentration parameter $c$) of $1200_{-100}^{+90}$ kpc ($c=13$) and $2000\pm100$ kpc ($c=2.8$) with confidence $4.6\,\sigma$ and $6.8\,\sigma$, respectively.

\begin{figure}
	\epsscale{1.0}
	\plotone{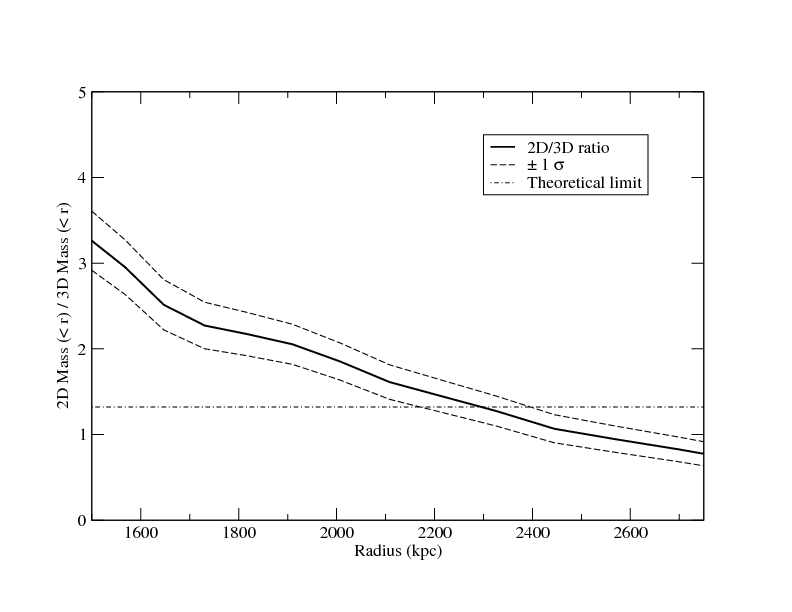}
	\caption{The solid line is the ratio of the 2D/3D mass of A1758N and the dashed lines show $\pm 1\,\sigma$.  The dot-dashed, horizontal line is the theoretical ratio ($\sim 1.32$) of the 2D/3D masses based on measurements of a collection of 30 X-ray luminous clusters from $z=0.15-0.3$ performed by \cite{okabe10}.  The background galaxies were put in equally spaced logarithmic bins based on radius.  Based on this 2D/3D ratio, the $r_{200}$ of A1758N is $2300_{-130}^{+100}$ kpc.}
	\label {fig:ratio2D3D}
\end{figure}

Given the problems with fitting the cores of the clusters to measure their mass, we also performed aperture mass densitometry \citep{clowe98}.  This routine works from the outer radii inward and gives a measurement of how much total mass there is along the line of sight out to any given radius:
\begin{align}
\nonumber \zeta_c(r_1) &= \bar{\kappa}(r\leq r_1) - \bar{\kappa}(r_2 \leq r \leq r_{max}) \\
&= 2 \int_{r_1}^{r_2} \langle \gamma_T \rangle\,d \ln r \nonumber \\
&~~~+ 2(1-r_2^2/r_{max}^2)^{-1} \int _{r_2} ^{r_{max}} \langle \gamma_T \rangle\, d \ln r,
\end{align}
\noindent 
where $\bar{\kappa}$ is the average $\kappa$ within a region and $\langle \gamma_T \rangle$ is the azimuthally averaged tangential shear.  This method assumes we are measuring the shear $\gamma$ and not the reduced shear $g$, but since we are only measuring this at large radius, the error from this approximation will be minimal.  This method, however, gives no indication of the virial mass of the cluster, so to determine the virial mass we used results by \cite{okabe10} from 30 X-ray luminous clusters from $z=0.15-0.3$ that gives $M_{200}$ based on the mass ratio $m_{2D}(<r)/m_{3D}(<r)$ of $\sim 1.32$ (see also \cite{metzler99}).  This ratio is due to the additional mass along the line of sight and because the cluster mass enclosed within a cylindrical volume is larger than the 3D NFW mass within the same radius.  Figure \ref{fig:ratio2D3D} shows where this mass ratio equals 1.32 $\pm 1\,\sigma$, where $m_{2D}$ is our aperture densitometry measurement at radius $r$ multiplied by $\Sigma_{crit}$ and $m_{3D}$ is the virial mass of an NFW profile with $r_{200}=r$.  The radii that fall within this range are $2170-2400$ kpc.  The ratio $m_{2D}/m_{3D}=1.32$ occurs at a radius of 2300 kpc and using this radius as an estimate of $r_{200}$ we obtain a 3D mass of $2.7 \pm 0.4 \times 10^{15} M_\odot$ for A1758N.  This is higher than the mass estimate of $1.6 \times 10^{15} M_\odot$ for A1758N given by \cite{david04} (based on X-ray temperature) for the northern clusters, but our measurements using this technique also includes a portion the mass of A1758S (2.0 Mpc away) within 2300 kpc.  The mass estimate of \cite{david04} is $1.0 \times 10^{15} M_{\odot}$ for the southern cluster.  At this distance, the amount of A1758S within 2300 kpc of the northern cluster is roughly 50\% ($\pm 10\%$ depending on concentration and morphology).  This reduces our 3D mass estimate of A1758N to $2.2 \pm 0.5 \times 10^{15} M_{\odot}$, which is still higher than, but in agreement at slightly higher than the $1\,\sigma$ level with, the mass estimate from \cite{david04}.

\subsection{Mass-to-Light Ratios}

The $M/L$ ratio in the region of the midpoint in A1758 is over 300 (see Table \ref{table:mass}), which is high compared to the values of our peaks $\sim200$.  Excess mass in this region could be caused by a self-interaction cross section for dark matter \citep{randall08}, some form of non-Newtonian gravity \citep{angus07}, or a higher concentration of galaxies compared to dark matter.  To test if the high $M/L$ ratio in the area between the clusters is significant, we simulated weak lensing observations for two NFW clusters the same size as the members of A1758 and performed the above analysis.  We found that the smoothing did not preferentially increase the mass in the region of the midpoint more than it did in any other direction.  

We also tested the catalog for the possibility of core galaxies being wrongly included with the background sample and decrease the reduced shear, and thus the mass, in that region.  We reduced the number of background galaxies in the cores by 10-30\% and the overall mass was reduced in the cores, but there was no change in the ratio of the $M/L$ ratios of the apertures.

We then compared the $M/L$ ratios of two other regions outside the merger that lie along the line of Peaks 1 and 2 and are the same distance away as is the midpoint, but in opposite directions (see Table \ref{table:mass}, ``Left'' and ``Right'').  These two regions were also found to have approximately the same $M/L$ ratio as the midpoint, which may indicate that either there is more dark matter or less red cluster light at this distance.

\subsection{A1758S}\label{subsec:A1758S}

\begin{figure*}
	\epsscale{1.0}
	\plotone{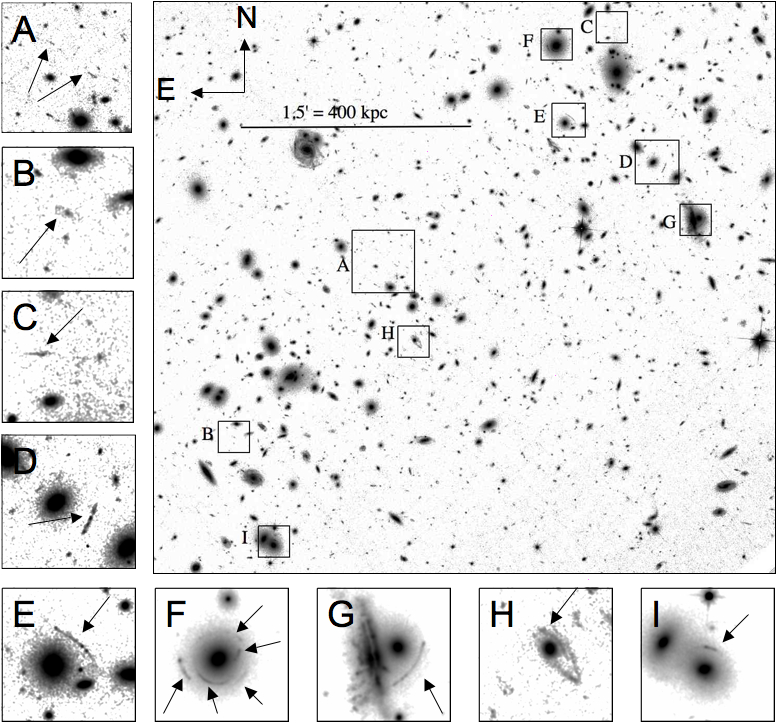}
	\caption{Hubble Space Telescope image of merging cluster A1758N.  Boxes indicate possible strong lensing candidates.  Regions $A-E$ match those of \citet{okabe08} and $F-I$ are possible new candidates.}
	\label {fig:strong}
\end{figure*}

We measured the mass of A1758S from the bootstrap resampling using aperture cutouts (see Section \ref {subsec:mass} and Table \ref{table:mass}).  The masses measured in the two southern clusters (Peaks 3 and 4, in units of $10^{13}\, M_\odot$) are $m_{Peak 3} = 5.72 \pm 1.1$, and $m_{Peak4} = 5.54 \pm 1.1$, or 5.2\,$\sigma$ and 5.0\,$\sigma$.

Peaks 3 and 4 were included in the bootstrap resampling to determine the errors on the centroid.  Of the $3\times10^5$ resamplings, Peaks 3 and 4 were found in the region of A1758S approximately 62\% of the time.  More specifically, Peak 3 was found in its own region 18\% of the time, but 45\% of the time it was found in the region of Peak 4.  The location of the centroid of Peak 4 was found in its own region 45\% of the time and it was located 16\% of the time in the region of Peak 3.  This is a measure of how little mass there is in A1758S and how difficult it was to distinguish between the two cluster members.

\cite{david04} find that the southern clusters are at their closest approach, but is disputed by \cite{haines09} who argue that a lack of star formation in the core galaxies of A1758S could be due to them being poorer clusters than A1758N.  We find that the $M/L$ ratio in the southern clusters (see Table \ref{table:mass}, Peaks 3 and 4) match those of the northern clusters, which does not agree with the description of \cite{haines09}.

The mass estimate of \cite{david04} for A1758S, based on X-ray gas mass, is $1.0\times10^{15} M_\odot$.  Compared to their estimate of A1758N ($1.6\times10^{15} M_\odot$), this gives a north/south mass ratio of 1.6.  Table \ref{table:mass} shows the core masses within identical apertures ($r=169.7$ kpc) of the northern clusters (Peaks 1 and 2) and southern clusters (Peaks 3 and 4).  The north/south mass ratio of our cluster cutouts is 1.53.

\subsection{Strong lensing candidates}\label{sec:stronglens}

Our image taken by the $Hubble$ $Space$ $Telescope$ ($HST$) Wide Field Planetary Camera 2 (see Figure \ref{fig:strong}) shows possible strong lensing candidates in the core region of A1758N.  Regions $A-E$ are strong lensing candidates from \citet{okabe08}.  In our own independent analysis we also find candidates $D-E$ (and possibly $C$) and discover regions $F-I$.  These are only candidates because we do not have spectroscopy or colors to determine more information about them.  Regions $F-I$ are not discernible in the Subaru images due to ground based seeing.  

In the Subaru $R_c$  images, Regions $A$ and $B$ appeared to be candidates, but at the seeing of the $Hubble$ image, $A$ is too faint to robustly confirm as an arc and $B$ appears to be a faint spiral galaxy whose length to width (L/W) ratio is $\sim3$ (see \cite{meneghetti01} for distinguishing arcs from edge-on spirals).  Region $C$ is a borderline strong lensing candidate with L/W $\sim5.6$, which could possibly have another image extending to the west in the image, but is too faint to be certain.  Regions $D$ and $E$ have L/W of $\sim8.4$ and $\sim12.5$, respectively.  Regions $F-I$ are newly discovered strong lensing candidates from the $HST$ image presented here.  Region $F$ is particularly interesting as we can see a nearly complete Einstein ring.  In addition, there seems to be a secondary source being lensed.  If such as system were confirmed, and the second source is at a different redshift, this would constrain cosmological parameters like $\Omega_{\rm m}$ and $\Omega_{\rm \Lambda}$ \citep{gavazzi08}.  Region $G$ is a good candidate, with L/W of $\sim17.4$, for being a strong arc and for being lensed by the edge-on spiral to the east.  Region $H$ appears to be a strong arc almost certainly influenced by the galaxy, but could be a spiral structure.  The candidate in Region $I$ has L/W of $\sim6.25$ of just the brightest region and may extend even further.  Most of these candidates appear to be primarily caused by individual cluster galaxy members, not the cluster core.  Only $C$, and possibly $D$, would provide any information about the cluster masses.  Without redshifts and multi-color images, we cannot use these to improve our mass measurements.  With high resolution, multi-color HST images (planned for Cycle 18, PI Clowe), faint sources that cannot be seen in the given shallow WFPC2 and ground-based Subaru observations will potentially be discovered.

\section{Summary}

A1758 is a region with two merging clusters in its northern component, A1758N, and two in its southern component, A1758S.  The merger A1758N shows a different geometry between its X-ray clouds and weak lensing peaks than previously published mergers.  Instead of having both X-ray clouds between the weak lensing peaks, just one X-ray cloud is between the weak lensing peaks and one X-ray cloud overlaps a weak lensing peak.  The location of the two mass peaks reconstructed from our weak lensing analysis are in good agreement with the positions of both BCGs of A1758N.

We calculated the error on the centroids of the two mass peaks in A1758N by bootstrapping the background galaxy catalog.  The two peaks were separated at the $2.5\,\sigma$ level compared to there being only one large mass with two noise bumps.  However, the reason that this procedure did not result in a more significant result is due to the noise in the mass map reconstructions rather than a lack of mass in either peak.  We found that 1-2\% of the time the centroid of one peak wandered across the midpoint (or to the location of the other peak) when its mass was low by $\sim 2\,\sigma$ and the mass around the midpoint was higher than it's average by $\sim 0.5\,\sigma$, creating a ramp for the peak finder to walk away from the original peak.

We performed a multiple cluster fitting of A1758 with two clusters in the north and a single cluster in the south (see Section \ref {subsec:virial} for details) and estimate $r_{200}$ (and concentration parameter $c$) for the two northern clusters to be $1200_{-100}^{+90}$ kpc ($c=13$) and $2000\pm100$ kpc ($c=2.8$) with $4.6\,\sigma$ and $6.8\,\sigma$ confidence, respectively.  However, because these parameters are quite degenerate, the fits can trade mass between the two peaks and these are likely not good estimates of the mass of each cluster.  We also estimated the virial radius of A1758N by using aperture mass densitometry, integrating from the outside in.  This method does not give the virial radius of the cluster so we used our 2D mass at each radius and the ratio given by \cite{okabe10} that compares $m_{2D}/m_{3D}=1.32$.  Our virial radius estimate for A1758N is $r_{200} = 2300_{-130}^{+100} $ kpc and, by the same method, our 3D mass estimate is $\sim2.2 \pm 0.5 \times10^{15} M_\odot$.  This is marginally higher than the $1.6 \times 10^{15} M_\odot$ estimate of \cite{david04}, which they estimated from X-ray temperatures. 



Our analysis does not reveal any new insights about the kinematics of clusters in A1758N or A1758S than have been previously presented \citep{david04,okabe08,haines09}.  The X-ray peak that overlaps the NW cluster member may simply be a line of sight effect, the merger may be in a later stage than the first pass, maybe the NW member is larger than the SE cluster and the merger had a large impact parameter, or perhaps the X-ray gas was not stripped from its host cluster due to less ram pressure.  Getting more redshifts for A1758 could improve our understanding of the dynamics of the northern and southern clusters, as would improving the S/N of the mass reconstruction using upcoming Hubble Space Telescope observations (Cycle 18, PI Clowe).

We detect seven strong lensing candidates in A1758N, three of which coincide with \cite{okabe08} and four new candidates.  With the improved seeing of the Hubble, candidate $A$ from \cite{okabe08} appears too faint to discern, candidate $B$ seems to be a faint spiral, while $C$ could be a candidate with a multiple image, but is too faint to discern.  Follow up observations with the Hubble Space Telescope are needed to confirm the candidacy of these objects and find multiply imaged source galaxies in the core regions and help determine the strong lensing radius of the clusters.  If candidates $C$ and $D$ are confirmed, they would help improve mass measurements of A1758N.  





\acknowledgments

Support for program HST-GO-11194.01-A was provided by NASA through a grant from the Space Telescope Science Institute, which is operated by the Association of Universities for Research in Astronomy, Inc., under NASA contract NAS 5-26555.  DC also acknowledges support from the Alfred P. Sloan Foundation.




{\it Facilities:} \facility{Subaru Suprime Cam}, \facility{Hubble Space Telescope}.




\end{document}